\documentstyle[12pt]{article}

\newcommand{\ba}{\begin{eqnarray}}
\newcommand{\ea}{\end{eqnarray}}
\input{psfig}
\def\ii{\'{\i}}
\begin{document}
\pagestyle{plain}

\title{Exactly Solvable Models of Baryon Structure \\ \mbox{} \\ 
Modelos Exactamente Solubles \\ de la Estructura Bari\'onica}
\author{R. Bijker\\
Instituto de Ciencias Nucleares,\\ 
Universidad Nacional Aut\'onoma de M\'exico,\\
A.P. 70-543, 04510 M\'exico, D.F., M\'exico 
\and
A. Leviatan\\
Racah Institute of Physics, The Hebrew University,\\
Jerusalem 91904, Israel}

\maketitle

\noindent
ABSTRACT: 
We present a qualitative analysis of the gross features of 
baryon spectroscopy (masses and form factors) in terms of 
various exactly solvable models. It is shown 
that a collective model, in which baryon resonances are 
interpreted as rotations and vibrations of an oblate symmetric top, 
provides a good starting point for a more detailed quantitative 
study. \\
\mbox{}\\
RESUMEN: 
Se presenta un an\'alisis cualitativo de las propiedades generales 
de la espectroscop\ii a bari\'onica (masas y factores de forma) 
en el contexto de varios modelos exactamente solubles. Se muestra 
que un modelo colectivo, en que se interpreta las resonancias 
bari\'onicas como rotaciones y vibraciones de un trompo oblato, 
provee un buen punto de partida para estudios m\'as detaillados.\\
\mbox{}

\clearpage

\section{Introduction} 

New dedicated experimental facilities ({\it e.g.} TJNAF, MAMI and 
ELSA) promise to produce new and more accurate data on the structure 
of the nucleon \cite{nstar}. The nucleon is not an elementary particle, 
but it is generally viewed as a confined system of quarks  
interacting via gluon exchange. Effective models of the nucleon 
and its excited states (or baryon resonances) 
are based on three constituent parts that carry the internal degrees 
of freedom of spin, flavor and color, but differ 
in their treatment of radial (or orbital) excitations. 

The purpose of this contribution is to discuss several exactly 
solvable models of baryon structure. Such models provide a set 
of closed analytic expressions for the mass spectrum, 
decay widths and selection rules that can be tested easily 
by experiment. We present a qualitative 
analysis of some salient features of baryon spectroscopy, such 
as the mass of the Roper resonance, the occurrence of linear Regge 
trajectories and the proton electric form factor, and discuss the 
merits and shortcomings of each scenario. 

\section{Algebraic models of the nucleon}

We consider baryons to be built of three constituent parts. The 
internal degrees of freedom of these three parts are taken to be:
flavor-triplet $u,d,s$ (for the light quark flavors), 
spin-doublet $S=1/2$, and color-triplet. The internal algebraic 
structure of the constituent parts consists of the usual spin-flavor 
($sf$) and color ($c$) algebras 
\ba 
{\cal G}_i &=& SU_{sf}(6) \otimes SU_{c}(3) ~.
\label{sfc}
\ea 
The relative motion of the three constituent parts can be described in
terms of Jacobi coordinates, $\vec{\rho}$ and $\vec{\lambda}$,
which in the case of three identical objects are 
\ba
\vec{\rho} &=& \frac{1}{\sqrt{2}} (\vec{r}_1 - \vec{r}_2) ~,
\nonumber\\
\vec{\lambda} &=& \frac{1}{\sqrt{6}} (\vec{r}_1 + \vec{r}_2 -2\vec{r}_3) ~. 
\label{jacobi}
\ea
Here $\vec{r}_1$, $\vec{r}_2$ and $\vec{r}_3$ are the coordinates of 
the three constituents. Instead of a formulation in terms of coordinates 
and momenta we use the method of bosonic quantization, 
in which we introduce a dipole boson with $L^P=1^-$ for each independent
relative coordinate, and an auxiliary scalar boson with $L^P=0^+$ 
\cite{BIL} 
\ba 
b^{\dagger}_{\rho,m} ~, \; b^{\dagger}_{\lambda,m} ~, \; 
s^{\dagger} \hspace{1cm} (m=-1,0,1) ~. \label{bb}
\ea
The scalar boson does not represent an independent degree of freedom, 
but is added under the restriction that the total number of bosons 
$N$ is conserved. This procedure leads to a compact spectrum generating 
algebra for the radial (or orbital) excitations 
\ba 
{\cal G}_r &=& U(7) ~.  \label{u7}
\ea 
For a system of interacting bosons the model space is spanned by the 
symmetric irreducible representation $[N]$ of $U(7)$. 
The value of $N$ determines the size of the model space. 

For nonstrange $qqq$ baryons, the mass operator 
has to be invariant under the permuation group $S_3$, {\it i.e.} 
under the interchange of any of the three constituent parts. 
This poses an additional constraint on the allowed interaction terms. 
The eigenvalues and corresponding eigenvectors can be 
obtained exactly by diagonalization in an appropriate basis. 
The wave functions have, by construction, good 
angular momentum $L$, parity $P$, and permutation symmetry $t$. 
The three symmetry classes of the $S_3$ permutation group are
characterized by the irreducible representations: 
$t=S$ for the one-dimensional symmetric representation, 
$t=A$ for the one-dimensional antisymmetric representation,  
and $t=M$ for the two-dimensional mixed symmetry representation. 

The $S_3$ invariant $U(7)$ mass operator has a rich group structure. 
It is of general interest to study limiting situations, in which 
the mass spectrum can be obtained in closed analytic form, that is 
to say, in terms of a mass formula. 
These special cases correspond to dynamical symmetries, and arise 
whenever the mass operator can be written in terms of Casimir invariants 
of a chain of subgroups of $U(7)$ only. 
Under the restriction that the eigenstates have good angular momentum, 
parity and permutation symmetry, there are several possibilities. 
Here we consider the chains 
\ba
U(7) \supset \left\{ \begin{array}{l} 
U(6) \supset \left\{ \begin{array}{l} {\cal SU}(3) \otimes SU(2) 
\supset {\cal SO}(3) \otimes SO(2) ~, \\ \mbox{} \\
SO(6) \supset SU(3) \otimes SO(2) 
\supset {\cal SO}(3) \otimes SO(2) ~, \end{array} \right. \\ \mbox{} \\
SO(7) \supset SO(6) \supset SU(3) \otimes SO(2) 
\supset {\cal SO}(3) \otimes SO(2) ~.
\end{array} \right. 
\label{lattice} 
\ea
The corresponding dynamical symmetries are referred to as the
$U(6) \supset {\cal SU}(3) \otimes SU(2)$ limit, the $U(6) \supset SO(6)$ 
limit and the $SO(7)$ limit, respectively. 
These chains have the direct product group ${\cal SO}(3) \otimes SO(2)$ 
in common, where ${\cal SO}(3)$ is the angular momentum group and 
$SO(2)$ is related to the permutation symmetry \cite{BIL,KM,Hey}. 

The mass operator depends both on the spatial and the internal degrees 
of freedom. We first discuss the contribution from the spatial part, 
which is obtained by expanding the mass-squared operator in terms of 
the generators of $U(7)$ \cite{BIL}. 

\subsection{The (an)harmonic oscillator}

\begin{figure}
\centering
\setlength{\unitlength}{1.0pt}
\begin{picture}(240,180)(0,0)
\thinlines
\put (  0,  0) {\line(1,0){240}}
\put (  0,180) {\line(1,0){240}}
\put (  0,  0) {\line(0,1){180}}
\put (240,  0) {\line(0,1){180}}
\thicklines
\put ( 70, 30) {\line(1,0){20}}
\put ( 70, 85) {\line(1,0){20}}
\put ( 70,150) {\line(1,0){20}}
\multiput ( 80,30)(0,5){25}{\circle*{0.1}}
\thinlines
\put ( 30, 25) {$n=0$}
\put ( 30, 80) {$n=1$}
\put ( 30,145) {$n=2$}
\put ( 95, 25) {$0^+_S$}
\put ( 95, 80) {$1^-_M$}
\put ( 95,145) {$2^+_S~, \, 2^+_M ~, \, 1^+_A ~, \, 0^+_S ~, \, 0^+_M$}
\put (180, 25) {$U(6)$}
\end{picture}
\vspace{15pt}
\caption[Anharmonic oscillator]
{Schematic spectrum of an anharmonic oscillator with $U(6)$ 
symmetry. The masses are calculated using Eq.~(\ref{ener1}) 
with $\epsilon_1>0$, $\epsilon_2>0$ and 
$\alpha=\kappa=\kappa^{\prime}=0$. The number of bosons is $N=2$.}
\label{anhosc}
\end{figure}
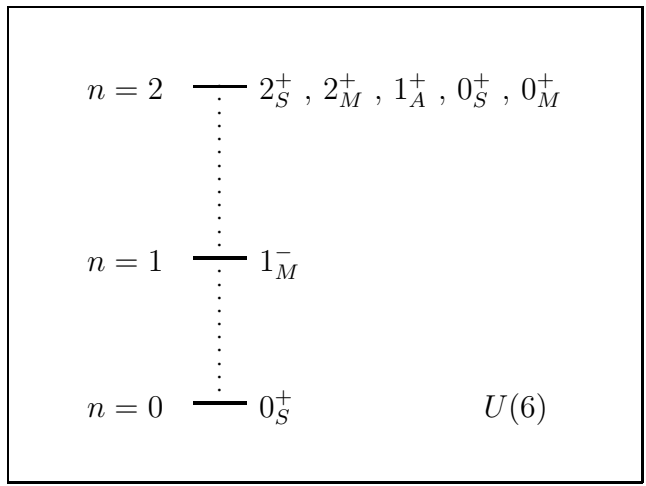

The first chain corresponds to the problem of three particles 
in a common harmonic oscillator potential \cite{KM}. It separates 
the behavior in three-dimensional coordinate space determined by 
${\cal SU}(3) \supset {\cal SO}(3)$, from that in the index space, 
given by $SU(2) \supset SO(2)$. The basis states are characterized by 
\ba
\left| \begin{array}{ccccccccccc}
U(7) &\supset& U(6) &\supset& {\cal SU}(3) &\otimes& SU(2) 
&\supset& {\cal SO}(3) &\otimes& SO(2) \\
N &,& n &,& && F &,& L &,& M_F \end{array} \right> ~. 
\label{ch1}
\ea
The irreducible representations of ${\cal SU}(3)$ are determined by 
$n$ and $F$, and hence do not provide independent labels. 
The allowed values of the quantum numbers can be obtained from the 
branching rules. Table~\ref{basis} shows the result for $N=2$. 
The mass spectrum in the $U(6) \supset {\cal SU}(3) \otimes SU(2)$ limit
is given by 
\ba
M^2(n,L,F,M_F) &=& M^2_0 + \epsilon_1 \, n + \epsilon_2 \, n(n+5) 
\nonumber\\
&&+ \alpha \, F(F+2) + \kappa \, L(L+1) + \kappa^{\prime} M_F^2 ~. 
\label{ener1}
\ea
In Fig.~\ref{anhosc} we show the structure of the spectrum 
of an anharmonic oscillator with $U(6)$ symmetry. 
The levels are grouped into oscillator shells characterized by $n$. 
The ground state has $n=0$ and $L^P_t=0^+_{S}$. The one-phonon 
multiplet $n=1$ has two degenerate states with $L^P=1^-$ which 
belong to the two-dimensional representation $M$ of the permutation group, 
and the two-phonon multiplet $n=2$ consists of the states 
$L^P_t=2^+_{S}$, $2^+_{M}$, $1^+_{A}$, $0^+_{S}$ and $0^+_{M}$. 
The $\epsilon_1$ term gives rise to a harmonic spectrum, whereas 
the $\epsilon_2$ term introduces anharmonicities. 
The splitting within an oscillator shell is determined by the last 
three terms of Eq.~(\ref{ener1}). In Fig.~\ref{ho1} we show the 
splitting for the $n=2$ multiplet. The terms proportional to 
$\alpha$ and $\kappa'$ separate the five states of the $n=2$ multiplet 
into two doublets with $L^P_t=2^+_{S}$, $0^+_{S}$ and 
$2^+_{M}$, $0^+_{M}$, respectively, and a $1^+_{A}$ singlet. 
Finally, the doublets can be split by adding the $L(L+1)$ term. 

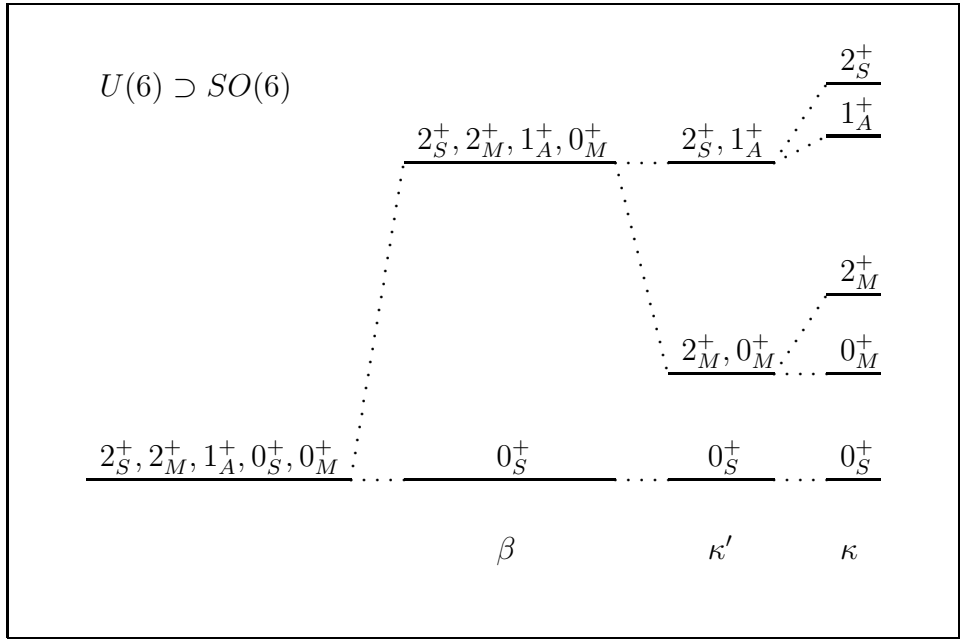
\begin{figure}
\centering
\setlength{\unitlength}{1.0pt}
\begin{picture}(360,240)(0,0)
\thinlines
\put (  0,  0) {\line(1,0){360}}
\put (  0,240) {\line(1,0){360}}
\put (  0,  0) {\line(0,1){240}}
\put (360,  0) {\line(0,1){240}}
\thicklines
\put ( 30, 60) {\line(1,0){100}}
\put (150, 60) {\line(1,0){80}}
\put (150,180) {\line(1,0){80}}
\put (250, 60) {\line(1,0){40}}
\put (250,100) {\line(1,0){40}}
\put (250,180) {\line(1,0){40}}
\put (310, 60) {\line(1,0){20}}
\put (310,100) {\line(1,0){20}}
\put (310,130) {\line(1,0){20}}
\put (310,190) {\line(1,0){20}}
\put (310,210) {\line(1,0){20}}
\multiput (130, 60)(5, 0){5}{\circle*{0.1}}
\multiput (130, 60)(0.82, 4.93){25}{\circle*{0.1}}
\multiput (230, 60)(5, 0){5}{\circle*{0.1}}
\multiput (230,180)(5, 0){5}{\circle*{0.1}}
\multiput (230,180)(1.21,-4.85){17}{\circle*{0.1}}
\multiput (290, 60)(5, 0){5}{\circle*{0.1}}
\multiput (290,100)(5, 0){5}{\circle*{0.1}}
\multiput (290,100)(2.77, 4.16){8}{\circle*{0.1}}
\multiput (290,180)(4.47, 2.24){5}{\circle*{0.1}}
\multiput (290,180)(2.77, 4.16){8}{\circle*{0.1}}
\thinlines
\put ( 35, 65) {$2^+_S,2^+_M,1^+_A,0^+_S,0^+_M$}
\put (185, 65) {$0^+_S$}
\put (155,185) {$2^+_S,2^+_M,1^+_A,0^+_M$}
\put (265, 65) {$0^+_S$}
\put (255,105) {$2^+_M,0^+_M$}
\put (255,185) {$2^+_S,1^+_A$}
\put (315, 65) {$0^+_S$}
\put (315,105) {$0^+_M$}
\put (315,135) {$2^+_M$}
\put (315,195) {$1^+_A$}
\put (315,215) {$2^+_S$}
\put (185, 30) {$\beta$}
\put (265, 30) {$\kappa'$}
\put (315, 30) {$\kappa$}
\put ( 35,205) {$U(6) \supset SO(6)$}
\end{picture}
\vspace{15pt}
\caption[The $U(6) \supset SO(6)$ limit]
{Splitting of the $n=2$ multiplet in the  
$U(6) \supset SO(6)$ limit. 
The masses are calculated using Eq.~(\ref{ener2}) 
by successively adding the terms 
with $\beta>0$, $\kappa'<0$ and $\kappa>0$.}
\label{ho2}
\end{figure}

Another classification scheme for the six-dimensional 
oscillator is provided by the second group chain of Eq.~(\ref{lattice}). 
The reduction $U(6) \supset SO(6) \supset SU(3) \otimes SO(2)$ 
has been studied in detail in \cite{Chacon}. Here it is embedded 
in $U(7)$. The corresponding basis states are characterized by the labels 
\ba
\left| \begin{array}{ccccccccccccc} 
U(7) &\supset& U(6) &\supset& SO(6) &\supset& SU(3) &\otimes& SO(2) 
&\supset& {\cal SO}(3) &\otimes& SO(2) \\ 
N &,& n &,& \sigma &,& && M_F  
&,& L && \end{array} \right> ~. \label{ch2} 
\ea
The irreducible representations of $SU(3)$ are determined by $\sigma$ 
and $M_F$, and hence do not provide independent labels. 
Table~\ref{basis} shows the classification scheme for $N=2$. 
The mass spectrum in the $U(6) \supset SO(6)$ limit is given by 
\ba
M^2(n,\sigma,L,M_F) &=& M^2_0 + \epsilon_1 \, n + \epsilon_2 \, n(n+5) 
\nonumber\\ 
&&+ \beta \, \sigma(\sigma+4) + \kappa \, L(L+1) + \kappa^{\prime} M_F^2 ~. 
\label{ener2}
\ea
Also in this case the eigenstates are grouped into oscillator shells 
according to Fig.~\ref{anhosc}. 
However, the splitting within an oscillator shell is different from 
that in the $U(6) \supset {\cal SU}(3) \otimes SU(2)$ limit. 
Fig.~\ref{ho2} shows that the terms proportional to 
$\beta$ and $\kappa'$ separate the five states of the $n=2$ multiplet 
into two doublets with $L^P_t=2^+_{S}$, $1^+_{A}$ and 
$2^+_{M}$, $0^+_{M}$, respectively, and a $0^+_{S}$ singlet. 
The difference between the two dynamical symmetries with $U(6)$ 
symmetry can be seen by comparing the position of the $0^+_S$  
and $1^+_A$ states in Figs.~\ref{ho1} and~\ref{ho2}.  

\subsection{The deformed oscillator}

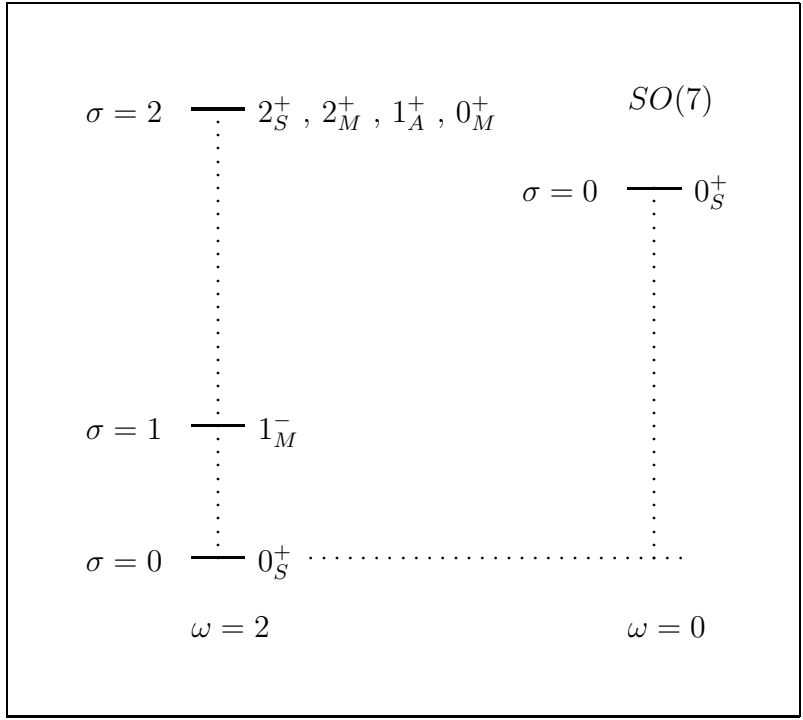
\begin{figure}
\centering
\setlength{\unitlength}{1.0pt}
\begin{picture}(300,270)(0,0)
\thinlines
\put (  0,  0) {\line(1,0){300}}
\put (  0,270) {\line(1,0){300}}
\put (  0,  0) {\line(0,1){270}}
\put (300,  0) {\line(0,1){270}}
\thicklines
\put ( 70, 60) {\line(1,0){20}}
\put ( 70,110) {\line(1,0){20}}
\put ( 70,230) {\line(1,0){20}}
\put (235,200) {\line(1,0){20}}
\multiput ( 80, 60)(0,5){35}{\circle*{0.1}}
\multiput (245, 60)(0,5){29}{\circle*{0.1}}
\multiput (115, 60)(5,0){29}{\circle*{0.1}}
\thinlines
\put ( 70, 30) {$\omega=2$}
\put (235, 30) {$\omega=0$}
\put ( 30, 55) {$\sigma=0$}
\put ( 30,105) {$\sigma=1$}
\put ( 30,225) {$\sigma=2$}
\put (195,195) {$\sigma=0$}
\put ( 95, 55) {$0^+_S$}
\put ( 95,105) {$1^-_M$}
\put ( 95,225) {$2^+_S ~, \, 2^+_M ~, \, 1^+_A ~, \, 0^+_M$}
\put (260,195) {$0^+_S$}
\put (235,230) {$SO(7)$}
\end{picture}
\vspace{15pt}
\caption[Deformed oscillator]
{Schematic spectrum of a deformed oscillator with $SO(7)$ 
symmetry. The masses are calculated using Eq.~(\ref{ener3}) 
with $A>0$, $\beta>0$ and $\kappa=\kappa^{\prime}=0$. 
The number of bosons is $N=2$.}
\label{defosc}
\end{figure}

The two group chains associated with the $U(7) \supset U(6)$ 
reduction correspond to a six-dimensional anharmonic oscillator, for 
which the total number of oscillator quanta $n$ is a good 
quantum number. This is no longer the case for the third 
dynamical symmetry of Eq.~(\ref{lattice}), for which the basis 
states are labeled by 
\ba
\left| \begin{array}{ccccccccccccc}
U(7) &\supset& SO(7) &\supset& SO(6) &\supset& SU(3) &\otimes& SO(2) 
&\supset& {\cal SO}(3) &\otimes& SO(2) \\
N &,& \omega &,& \sigma &,& && M_F 
&,& L && \end{array} \right> ~. \label{ch3}
\ea 
The classification scheme of the basis states for $N=2$ is given in 
Table~\ref{basis}. The $SO(7)$ limit corresponds to a deformed 
oscillator and its mass spectrum is given by
\ba
M^2(\omega,\sigma,L,M_F) &=& M^2_0 + A \, (N-\omega)(N+\omega+5) 
\nonumber\\ 
&& + \beta \, \sigma(\sigma+4) + \kappa \, L(L+1) + \kappa^{\prime} M_F^2 ~. 
\label{ener3}
\ea
In Fig.~\ref{defosc} we show the mass spectrum of the deformed 
oscillator with $SO(7)$ symmetry. The states 
are now ordered in bands characterized by 
$\omega$, rather than in harmonic oscillator shells, as in the 
previous two examples. Although the size of the model space,  
and hence the total number of states, is the same as for the  
harmonic oscillator, the ordering and classification 
of the states is different. In the $U(6)$ limit, the states with 
$L^P_t=2^+_{S}$, $2^+_{M}$, $1^+_{A}$, $0^+_{S}$ and $0^+_{M}$ 
form the two-phonon multiplet, whereas in the $SO(7)$ limit 
the $0^+_S$ state is a vibrational excitation whose 
energy is determined by the value of $A$, and the 
$2^+_{S}$, $2^+_{M}$, $1^+_{A}$, $0^+_{M}$ states are rotations 
which belong to the ground state band with $\omega=N$ and 
whose excitation energies depend on $\beta$, $\kappa$ and $\kappa'$. 

\subsection{The oblate symmetric top}

\begin{figure}
\centering
\setlength{\unitlength}{0.9pt}
\begin{picture}(460,305)(0,0)
\thinlines
\put (  0,  0) {\line(1,0){460}}
\put (  0,305) {\line(1,0){460}}
\put (  0,  0) {\line(0,1){305}}
\put (460,  0) {\line(0,1){305}}
\thicklines
\put ( 30, 60) {\line(1,0){20}}
\put ( 30,100) {\line(1,0){20}}
\put ( 30,140) {\line(1,0){20}}
\put ( 70,100) {\line(1,0){20}}
\put ( 70,140) {\line(1,0){20}}
\put (110,140) {\line(1,0){20}}
\multiput ( 40,160)(0,5){5}{\circle*{0.1}}
\multiput ( 70, 60)(5,0){68}{\circle*{0.1}}
\thinlines
\put ( 40, 25) {$(v_1,v_2)=(0,0)$}
\put ( 52, 55) {$0^+_S$}
\put ( 52, 95) {$1^+_A$}
\put ( 92, 95) {$1^-_M$}
\put ( 52,135) {$2^+_S$}
\put ( 92,135) {$2^-_M$}
\put (132,135) {$2^+_M$}
\thicklines
\put (160, 90) {\line(1,0){20}}
\put (160,130) {\line(1,0){20}}
\put (160,170) {\line(1,0){20}}
\put (200,130) {\line(1,0){20}}
\put (200,170) {\line(1,0){20}}
\put (240,170) {\line(1,0){20}}
\multiput (170,190)(0,5){5}{\circle*{0.1}}
\multiput (170, 60)(0,5){7}{\circle*{0.1}}
\thinlines
\put (170, 25) {$(v_1,v_2)=(1,0)$}
\put (182, 85) {$0^+_S$}
\put (182,125) {$1^+_A$}
\put (222,125) {$1^-_M$}
\put (182,165) {$2^+_S$}
\put (222,165) {$2^-_M$}
\put (262,165) {$2^+_M$}
\thicklines
\put (290,150) {\line(1,0){20}}
\put (290,190) {\line(1,0){20}}
\put (290,230) {\line(1,0){20}}
\put (330,190) {\line(1,0){20}}
\put (330,230) {\line(1,0){20}}
\put (385,230) {\line(1,0){20}}
\multiput (300,250)(0,5){5}{\circle*{0.1}}
\multiput (300, 60)(0,5){19}{\circle*{0.1}}
\thinlines
\put (300, 25) {$(v_1,v_2)=(0,1)$}
\put (312,145) {$0^+_M$}
\put (312,185) {$1^+_M$}
\put (352,185) {$1^-_{SAM}$}
\put (312,225) {$2^+_M$}
\put (352,225) {$2^-_{SAM}$}
\put (407,225) {$2^+_{SAM}$}
\end{picture}
\vspace{1cm}
\caption[Oblate top]
{Schematic spectrum of an oblate symmetric top. 
The masses are calculated using Eq.~(\ref{evib}) 
with $\kappa_1>0$, $\kappa_2>0$ and an additional  
term linear in $L$.}
\label{top}
\end{figure}
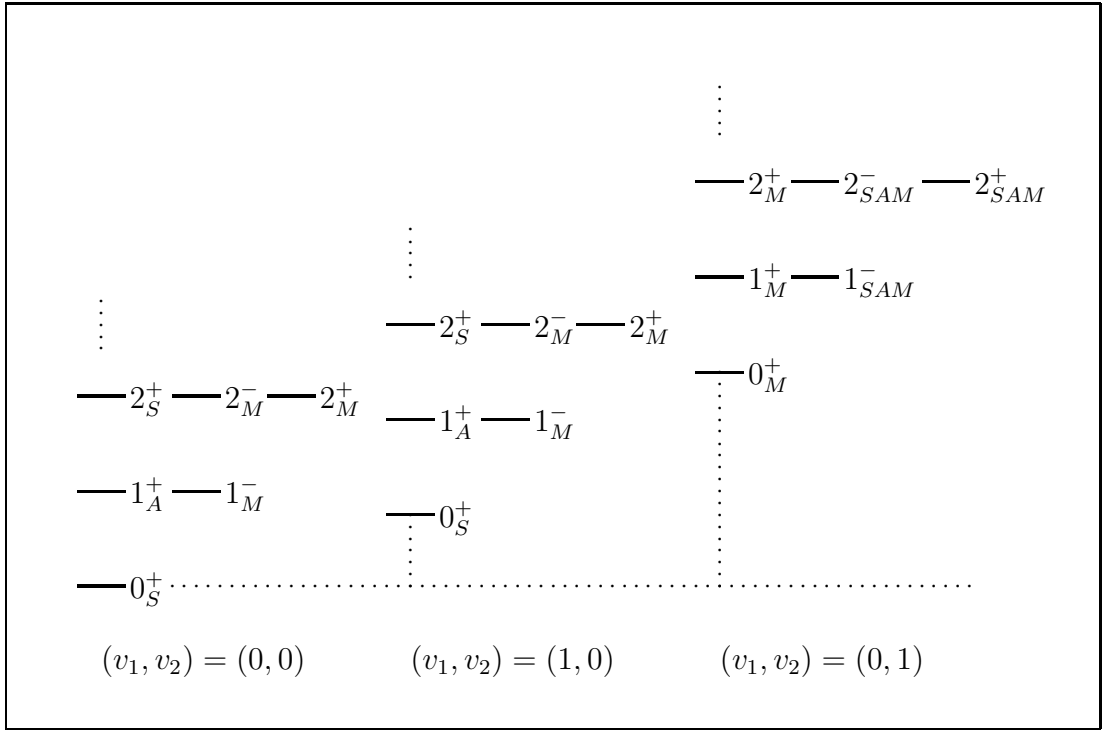

Another interesting case, that does not correspond 
to a dynamical symmetry, is provided by the mass operator 
\ba
M^2 &=& \xi_1 \, ( R^2 \, s^{\dagger} s^{\dagger}
- b^{\dagger}_{\rho} \cdot b^{\dagger}_{\rho}
- b^{\dagger}_{\lambda} \cdot b^{\dagger}_{\lambda} ) \,
( R^2 \, \tilde{s} \tilde{s} - \tilde{b}_{\rho} \cdot \tilde{b}_{\rho}
- \tilde{b}_{\lambda} \cdot \tilde{b}_{\lambda} )
\nonumber\\
&& + \xi_2 \, \left[ ( b^{\dagger}_{\rho} \cdot b^{\dagger}_{\rho}
- b^{\dagger}_{\lambda} \cdot b^{\dagger}_{\lambda} ) \,
( \tilde{b}_{\rho} \cdot \tilde{b}_{\rho}
- \tilde{b}_{\lambda} \cdot \tilde{b}_{\lambda} )
+ 4 \, ( b^{\dagger}_{\rho} \cdot b^{\dagger}_{\lambda} ) \,
( \tilde{b}_{\lambda} \cdot \tilde{b}_{\rho} ) \right] ~, 
\label{oblate}
\ea
with $\tilde{s}=s$~, $\tilde{p}_{\rho,m}=(-1)^{1-m} p_{\rho,-m}$ and 
$\tilde{p}_{\lambda,m}=(-1)^{1-m} p_{\lambda,-m}$~. 
For $R^2=0$ the mass operator of 
Eq.~(\ref{oblate}) has $U(7) \supset U(6)$ symmetry and corresponds
to an anharmonic vibrator, whereas for $R^2=1$ and $\xi_2=0$ it has 
$U(7) \supset SO(7)$ symmetry and corresponds to a deformed 
oscillator. The general case with $R^2 \neq 0$ and $\xi_1$, $\xi_2>0$ 
corresponds to an oblate symmetric top \cite{BIL}. 
In a normal mode analysis the mass operator of Eq.~(\ref{oblate}) 
reduces to leading order in $N$ to a harmonic form,  
and its spectrum is given by \cite{BIL} 
\ba
M^2(v_1,v_2) &=& \kappa_1 \, v_1 + \kappa_2 \, v_2 ~, 
\label{evib}
\ea
with $\kappa_1 = \xi_1 \, 4 N R^2$ and 
$\kappa_2 = \xi_2 \, 4 N R^2/(1+R^2)$. 
Here $v_1$ and $v_2$ represent vibrational quantum numbers. 
In Fig.~\ref{top} we show a schematic spectrum of an oblate 
symmetric top. In anticipation of the application to the 
mass spectrum of nonstrange baryon resonances we have added 
a term linear in the angular momentum $L$. 
The spectrum consists of a series of vibrational excitations 
characterized by the labels $(v_1,v_2)$, and a tower of 
rotational excitations built on top of each vibration. 

\subsection{The hypercoulomb potential}

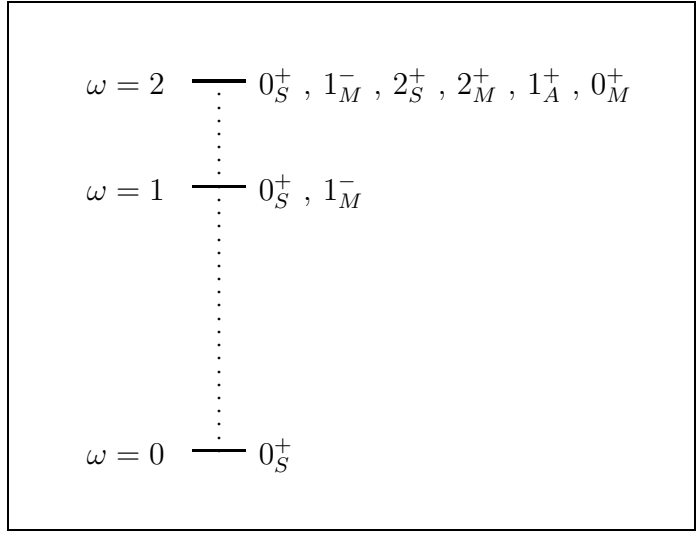
\begin{figure}
\centering
\setlength{\unitlength}{1.0pt}
\begin{picture}(260,200)(0,0)
\thinlines
\put (  0,  0) {\line(1,0){260}}
\put (  0,200) {\line(1,0){260}}
\put (  0,  0) {\line(0,1){200}}
\put (260,  0) {\line(0,1){200}}
\thicklines
\put ( 70, 30) {\line(1,0){20}}
\put ( 70,130) {\line(1,0){20}}
\put ( 70,170) {\line(1,0){20}}
\multiput ( 80,30)(0,5){29}{\circle*{0.1}}
\thinlines
\put ( 30, 25) {$\omega=0$}
\put ( 30,125) {$\omega=1$}
\put ( 30,165) {$\omega=2$}
\put ( 95, 25) {$0^+_S$}
\put ( 95,125) {$0^+_S~, \, 1^-_M$}
\put ( 95,165) {$0^+_S~, \, 1^-_M~, \, 
2^+_S~, \, 2^+_M ~, \, 1^+_A ~, \, 0^+_M$}
\end{picture}
\vspace{15pt}
\caption[Hypercoulomb potential]
{Schematic spectrum of a hypercoulomb potential model. 
The masses are calculated using Eq.~(\ref{ehc}) 
with $\tau^2 \mu>0$.}
\label{coulomb}
\end{figure}

Another exactly solvable model is provided by the 
hypercoulomb potential in six dimensions 
\ba
H &=& \frac{1}{2\mu} \, (\vec{p}_{\rho}^{\,2} 
+ \vec{p}_{\lambda}^{\,2}) - \frac{\tau}{x} ~, 
\label{hcpot}
\ea
with $x = \sqrt{ \vec{\rho}^{\,2} + \vec{\lambda}^{\,2}}$.  
This model can be solved in closed form by introducing hyperspherical 
coordinates \cite{Leal,SIG} or, equivalently, by using the properties of 
the noncompact dynamical group \cite{Barut,BIS} 
\ba 
{\cal G}_r &=& SO(7,2) ~.  \label{hc}
\ea 
The energy spectrum is given by  
\ba 
E(\omega) &=& - \frac{\tau^2 \mu}{2(\omega+\frac{5}{2})^2} ~. 
\label{ehc}
\ea 
The eigenstates are grouped into multiplets labeled by $\omega$. 
The degenerate states can be classified according to the 
symmetric representations of $SO(7)$ and its subgroups 
(see Table~\ref{basis}). 
In Fig.~\ref{coulomb} we show the corresponding spectrum. 
It is important to note that, although the eigenvalues can 
be labeled by $\omega$, the eigenfunctions of Eq.~(\ref{hcpot}) 
are not $SO(7)$ wave functions \cite{BIS}. 

\section{Mass spectrum}

Here we consider the nonstrange baryon resonances belonging to the 
nucleon (isospin $I=1/2$) and the delta (isospin $I=3/2$) family. 
The full algebraic structure is obtained by combining the spatial 
part ${\cal G}_r$ of Eq.~(\ref{u7}) or (\ref{hc}) with the internal 
spin-flavor-color part ${\cal G}_i$ of Eq.~(\ref{sfc})
\ba
{\cal G} &=& {\cal G}_r \otimes SU_{sf}(6) \otimes SU_c(3) ~.
\ea
The spatial part of the baryon wave function has to be combined 
with the spin-flavor and color part, in such a way that the total 
wave function is antisymmetric. Since the color part of the wave 
function is antisymmetric (color singlet), the remaining part 
(spatial plus spin-flavor) has to be symmetric. For nonstrange 
resonances which have three identical constituent parts this means 
that the symmetry of the spatial wave function under $S_3$ is 
the same as that of the spin-flavor part. Therefore, one can use the 
representations of either $S_3$ or $SU_{sf}(6)$ to label the states. 
The subsequent decomposition of representations of 
$SU_{sf}(6)$ into those of $SU_{f}(3) \otimes SU_{s}(2)$
is the standard one
\ba
S \;\leftrightarrow\; [56] &\supset& ^{2}8 \, \oplus \, ^{4}10 ~,
\nonumber\\
M \;\leftrightarrow\; [70] &\supset& 
^{2}8 \, \oplus \, ^{4}8 \, \oplus \, ^{2}10 \, \oplus \, ^{2}1 ~,
\nonumber\\
A \;\leftrightarrow\; [20] &\supset& ^{2}8 \, \oplus \, ^{4}1 ~.
\ea
The representations of the spin-flavor groups $SU_{sf}(6)$, 
$SU_f(3)$ and $SU_s(2)$ are denoted by their dimensions. 
The total baryon wave function is expressed as 
\ba
\left| \Psi \right> &=& \left| \, ^{2S+1}\mbox{dim}\{SU_f(3)\}_J \, 
[\mbox{dim}\{SU_{sf}(6)\},L^P] \, \right> ~,
\label{baryonwf}
\ea
where $S$ and $J$ are the spin and total angular momentum 
$\vec{J}=\vec{L}+\vec{S}$. As an example, the nucleon wave 
function is given by 
\ba
\left| \Psi \right> &=& \left| \, ^{2}8_{1/2} \, [56,0^+] \, \right> ~. 
\label{nucleonwf}
\ea

In the previous section we have discussed four exactly solvable 
models to describe the radial (or orbital) excitations of baryons. 
The mass spectrum of nonstrange baryons is characterized by 
the lowlying N(1440) resonance with $J^P=1/2^+$ (the socalled Roper 
resonance), whose mass is smaller than that of the first excited 
negative parity resonances, and the occurrence of linear Regge 
trajectories. These two features make an interpretation in terms 
of the harmonic vibrator, the deformed oscillator or the 
hypercoulomb potential model difficult. 

The Roper resonance has 
the same quantum numbers as the nucleon of Eq.~(\ref{nucleonwf}), 
but is associated with the first excited $L_t^P=0^+_S$ state.
In the harmonic oscillator the first excited $L_t^P=0^+_S$ state 
belongs to the $n=2$ positive parity multiplet which lies above 
the first excited negative parity state with $n=1$ (see 
Fig.~\ref{anhosc}), whereas the data shows the opposite. 
In the hypercoulomb 
potential model the Roper resonance occurs at the same mass as the 
negative parity resonances, whereas in the deformed oscillator and the  
oblate top the Roper resonance is a vibrational excitation whose 
mass is independent of that of the negative parity states 
which are interpreted as rotational excitations. 

\begin{figure}
\centerline{\hbox{
\psfig{figure=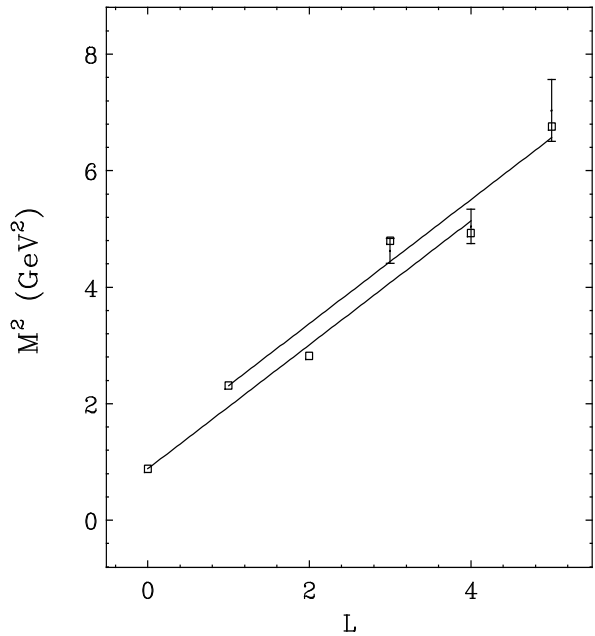,height=0.75\textwidth,width=1.0\textwidth,angle=180} }}
\caption[Regge trajectories]
{Regge trajectories for the positive parity resonances 
$| ^{2}8_{J=L+1/2} \, [56,L^+] \rangle$ with $L=0,2,4$ and 
the negative parity resonances  
$| ^{2}8_{J=L+1/2} \, [70,L^-] \rangle$ with $L=1,3,5$. 
The lines represent the result for the oblate symmetric top.}
\label{Regge}
\end{figure}

Furthermore, the data show that the mass-squared of the resonances 
depends linearly on the orbital angular momentum $M^2 \propto L$. 
The resonances belonging to such a Regge trajectory have the same 
quantum numbers with the exception of $L$. In Fig.~\ref{Regge} we 
show the trajectories for the positive parity resonances 
$| ^{2}8_{J=L+1/2} \, [56,L^+] \rangle$ with $L=0,2,4$ and 
for the negative parity resonances  
$| ^{2}8_{J=L+1/2} \, [70,L^-] \rangle$ with $L=1,3,5$. 
The splitting of the rotational states in the harmonic oscillator 
and the hypercoulomb potential model is hard to reconcile with linear 
Regge trajectories. 

In the oblate top and the deformed oscillator it is straightforward 
to reproduce the relative mass of the Roper resonance and the 
occurrence of linear Regge trajectories. The main difference between 
these two models is the interpretation of the first excited $0^+_M$ 
state. In the deformed oscillator it is a rotational excitation that 
belongs to the ground band $(\omega=N)$, whereas in the oblate top 
it is a fundamental vibration. In Tables~\ref{nucleon} and \ref{delta} 
we present an analysis of nonstrange baryon resonances in terms of 
vibrations and rotations of an oblate symmetric top. 
A simultaneous fit of the 25 well-established (3 and 4 star) nucleon 
and delta resonances of \cite{PDG} gives a r.m.s. deviation of 39 MeV 
\cite{BIL}. The 1 and 2 star resonances are not very well established 
experimentally. For some of these resonances we have indicated several 
possible assignments. In addition to the resonances presented in 
Tables~\ref{nucleon} and \ref{delta} there are many more states 
calculated than have been observed so far, especially in the nucleon 
sector. The lowest socalled 
`missing' resonances correspond to the unnatural parity states 
with $L^P=1^+$, $2^-$, which are decoupled both in electromagnetic 
and strong decays, and hence very difficult to observe. 

\section{Form factors}

In addition to the mass spectrum it is important to investigate 
the decay channels of baryon resonances. 
The generic form of a (partial) decay width is 
\ba
\Gamma &=& \rho \, \chi \, | F(k) |^2 ~,
\ea
where $\rho$ is a phase space factor, $\chi$ contains the 
spin-flavor dependence and $F(k)$ is the geometric form factor 
that depends on the momentum transfer. 
All models discussed in this article have the same spin-flavor 
structure, but differ in the treatment of the radial excitations, 
and hence in the geometric form factor. Helicity asymmetries 
of baryon resonances 
$(|A_{1/2}|^2 - |A_{3/2}|^2) / (|A_{1/2}|^2 + |A_{3/2}|^2)$ 
only depend on the spin-flavor part and cannot be used to 
distinguish between the different models of baryon structure, 
since in the ratio of the difference and the sum of the 
helicity $1/2$ and $3/2$ amplitudes the phase space factor 
and the geometric form factor cancel. 

However, electromagnetic form factors which can be measured in 
electroproduction of baryon resonances do provide a good test 
of various models of baryon structure. Here we use the 
proton electric form factor $G_E^p$ as an example. 
It can be shown that $G_E^p$ is proportional to the radial 
matrix element 
\ba
G_E^p(k) &=& \langle \Psi_p | \, 
\mbox{e}^{-ik \sqrt{\frac{2}{3}} \lambda_z} \, | \Psi_p \rangle ~.
\ea
In the hypercoulomb potential model this matrix element can be 
derived in closed form either in coordinate space \cite{Leal,SIG} 
or algebraically \cite{BIS} 
\ba
G_E^p(k) \;=\; \frac{1}{(1+k^2 b^2)^{7/2}} 
\;=\; 1 - \frac{7}{2} k^2 b^2 + \ldots ~, 
\label{geph}
\ea
with $b^2 = 25/24 \tau^2 \mu^2$. In principle, 
the coefficient $b$ can be determined by the charge 
radius of the proton $b^2 = \langle r^2 \rangle_p/21$. 

In $U(7)$ the Jacobi coordinate $\lambda_z$ is expressed in terms 
of algebraic operators as \cite{BIL,emff} 
\ba
\sqrt{\frac{2}{3}} \lambda_z &\rightarrow& \beta \, \hat D_{\lambda,z}/X_D ~,
\ea
where the dipole operator $\hat D_{\lambda,m}= 
(b^{\dagger}_{\lambda} \tilde{s} - s^{\dagger} \tilde{b}_{\lambda})^{(1)}_m$  
is a generator of $U(7)$ that transforms as a vector under rotations, 
is even under time reversal, and transforms under permutations 
as the Jacobi coordinate $\lambda_m$. The coefficient $X_D$ is a 
normalization factor and $\beta$ represents the scale of the coordinate. 
The proton radial wave function $| \Psi_p \rangle$ depends on the model 
that is used to describe the orbital excitations. 
For the three limiting cases of $U(7)$ that were discussed above 
we obtain 
\ba
G_E^p(k) &=& \left\{ \begin{array}{ll} 
\mbox{e}^{-k^2 \beta^2 /6} \;=\; 1 - k^2 \beta^2/6 + \ldots & 
\mbox{Harmonic oscillator} \\ \mbox{} \\
4 J_2(k \beta \sqrt{2})/k^2 \beta^2 \;=\; 1 - k^2 \beta^2/6 + \ldots & 
\mbox{Deformed oscillator} \\ \mbox{} \\
j_0(k \beta) \;=\; 1 - k^2 \beta^2/6 + \ldots & 
\mbox{Oblate top} 
\end{array} \right. \label{gep0}
\ea
The scale parameter $\beta$ can be determined by the charge 
radius of the proton $\beta^2 = \langle r^2 \rangle_p$. 

\begin{figure}
\centerline{\hbox{
\psfig{figure=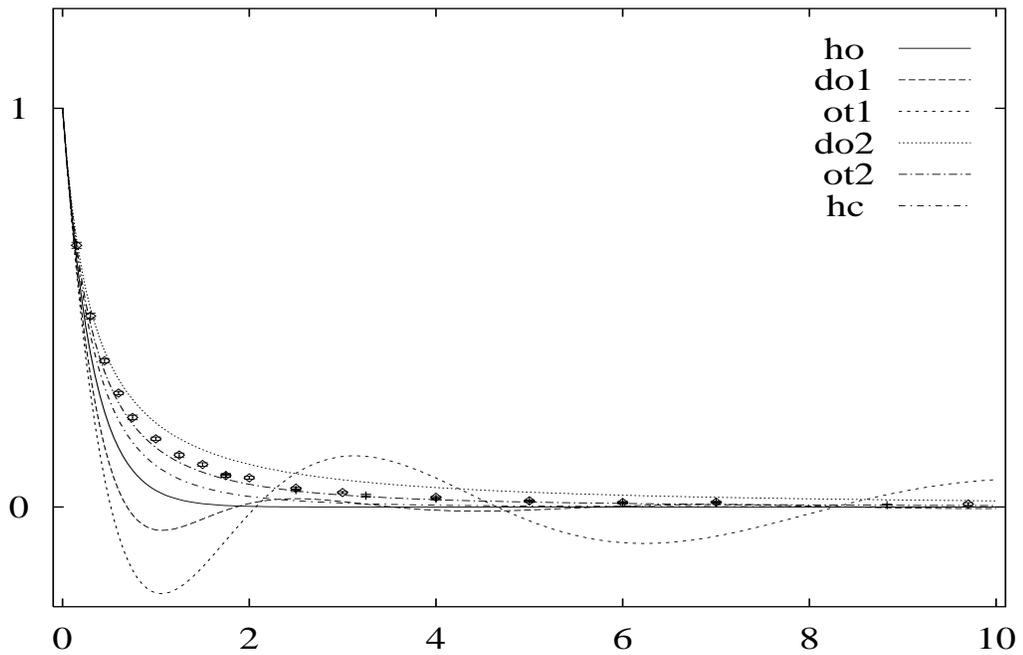,height=0.55\textwidth,width=1.0\textwidth} }}
\caption[Proton electric form factor: $G_E^p$]
{Comparison between the experimental and calculated proton electric form 
factor $G_E^p$ as a function of $Q^2=k^2$.
The curves labeled by ho, do1, ot1, 
do2, ot2 and hc are calculated with Eqs.~(\ref{gep0}), (\ref{gep2}), 
(\ref{gep1}) and (\ref{geph}), respectively. 
The experimental data are taken from \cite{Walker} 
and \cite{Andivahis}.}
\label{gep}
\end{figure}

\begin{figure}
\centerline{\hbox{
\psfig{figure=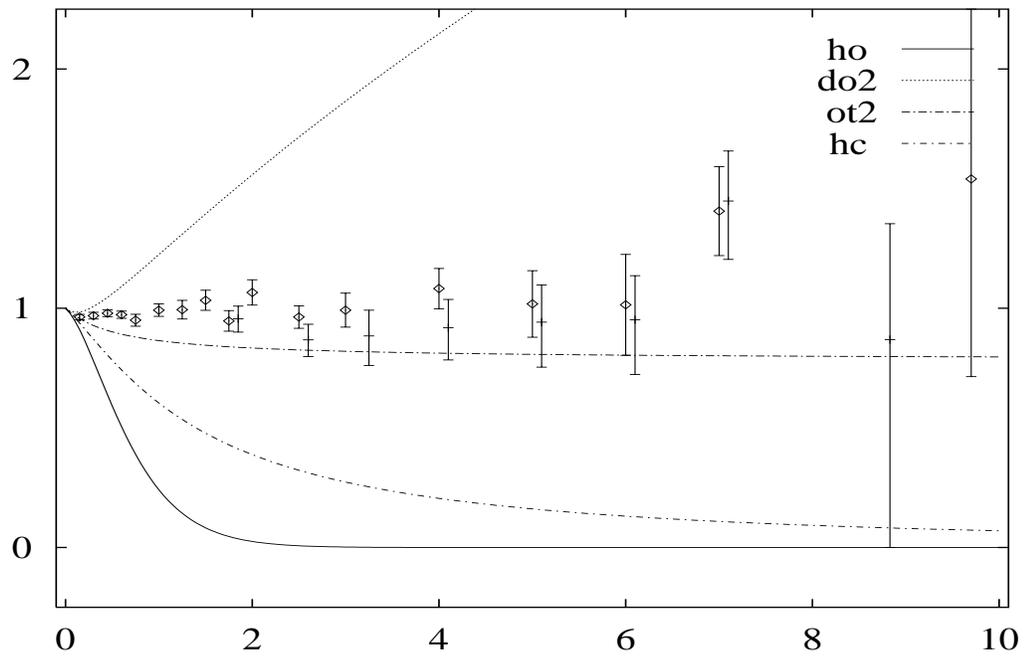,height=0.55\textwidth,width=1.0\textwidth} }}
\caption[Proton electric form factor: $G_E^p/F_D$ ]
{As Fig.~\ref{gep}, but divided by the dipole form factor 
$1/(1+k^2/0.71)^2$.}
\label{gepfd}
\end{figure}

In a collective model of baryons the nucleon has a specified distribution 
of charge and magnetization \cite{BIL,emff}
\ba
g(\beta) = \beta^2 \, \mbox{e}^{-\beta/a} / 2a^3 ~, 
\label{gbeta} 
\ea
where $a$ is a scale parameter. The ansatz of Eq.~(\ref{gbeta}) was  
made to obtain the dipole form for the oblate top 
\ba
G_E^p(k) \;=\; \int_0^{\infty} d\beta \, g(\beta) \, j_0(k \beta) \;=\; 
\frac{1}{(1+k^2 a^2)^2} \;=\; 1 - 2 k^2 a^2 + \ldots ~. 
\label{gep1}
\ea 
For the deformed oscillator we find 
\ba
G_E^p(k) \;=\; \frac{(\sqrt{1 + 2 k^2 a^2} - 1)^2}
{k^4 a^4 \sqrt{1 + 2 k^2 a^2}} \;=\; 1 - 2 k^2 a^2 + \ldots ~. 
\label{gep2}
\ea
Again the scale parameter $a$ can be determined by the charge 
radius of the proton $a^2 = \langle r^2 \rangle_p/12$. For small 
momentum transfer the behavior of the proton electric form 
factor is the same for all cases considered, since it is 
determined by the proton charge radius. However, for large 
momentum transfer the behavior is very different. Eq.~(\ref{gep0}) 
shows that, whereas for the harmonic oscillator $G_E^p$ drops 
exponentially, for the deformed oscillator and the oblate top 
it exhibits an oscillatory behavior. In a collective model of the 
nucleon introduced in Eq.~(\ref{gbeta}) the proton electric form 
factor drops as $k^{-4}$ for the oblate top (by construction), 
and as $k^{-3}$ for the deformed oscillator. 
For the hypercoulomb potential model $G_E^p$ drops as $k^{-7}$.  
In Fig.~\ref{gep} we show a comparison with the data. 
In Fig.~\ref{gepfd} we have divided $G_E^p$ by a dipole form factor 
$1/(1+k^2/0.71)^2$ to show more clearly the differences between 
the various model calculations. 

\section{Summary and conclusions}

In summary, we have presented a qualitative analysis of 
the mass spectrum and form factors 
of nonstrange baryons in the context of several exactly solvable 
models. These models share a common spin-flavor structure, but 
differ in their treatment of radial excitations.  
We first discussed a $U(7)$ interacting boson model of baryons for the 
spatial degrees of freedom \cite{BIL}. It was shown that this 
model unifies various exactly solvable models and, as such, 
provides a general framework to study the properties of baryon 
resonances in a transparent and systematic way. 
For nonstrange baryon resonances it contains 
the (an)harmonic vibrator, the deformed oscillator and 
the oblate symmetric top as special cases. Another solvable model 
that we have considered is the hypercoulomb potential 
model. Although the hypercoulomb potential is not confining, it is 
nevertheless worthwhile to examine its main features. Confining 
terms can added later and studied numerically \cite{SIG}. 

It was shown that a collective model, in which baryon resonances are 
interpreted as rotations and vibrations of an oblate symmetric top, 
can account for the mass of the Roper resonance, the linear Regge 
trajectories and the proton electric form factor. Hence this scenario 
provides a good starting point for a more detailed quantitative 
study of baryons. 

\section*{Acknowledgements}

It is a pleasure to thank F. Iachello and E. Santopinto for 
interesting discussions. 
This work is supported in part by DGAPA-UNAM under project IN101997,
and by grant No. 94-00059 from the United States-Israel Binational
Science Foundation.

\clearpage

\begin{table}
\centering
\caption[Basis states]{Classification scheme of the basis states in 
(a) the $U(6) \supset {\cal SU}(3) \otimes SU(2)$ limit, 
(b) the $U(6) \supset SO(6)$ limit, and 
(c) the $SO(7)$ limit. The number of bosons is $N=2$.}
\label{basis} 
\vspace{15pt}
\begin{tabular}{ccrl|ccrl|ccrl} 
\hline
& & & & & & & & & & & \\
\multicolumn{4}{c|}{(a)} &
\multicolumn{4}{c|}{(b)} &
\multicolumn{4}{c}{(c)} \\
$n$ & $F$ & $M_F$ & $L^P_t$ & 
$n$ & $\sigma$ & $M_F$ & $L^P_t$ & 
$\omega$ & $\sigma$ & $M_F$ & $L^P_t$ \\
& & & & & & & & & & & \\
\hline
& & & & & & & & & & & \\
0 & 0 &       0 & $0^+_S$ &
0 & 0 &       0 & $0^+_S$ &
2 & 0 &       0 & $0^+_S$ \\
1 & 1 & $\pm 1$ & $1^-_M$ &
1 & 1 & $\pm 1$ & $1^-_M$ &
  & 1 & $\pm 1$ & $1^-_M$ \\
2 & 2 & $\pm 2$ & $0^+_M,2^+_M$ &
2 & 2 & $\pm 2$ & $0^+_M,2^+_M$ &
  & 2 & $\pm 2$ & $0^+_M,2^+_M$ \\
  &   &       0 & $0^+_S,2^+_S$ &
  &   &       0 & $1^+_A,2^+_S$ &
  &   &       0 & $1^+_A,2^+_S$ \\
  & 0 &       0 & $1^+_A$ &
  & 0 &       0 & $0^+_S$ &
0 & 0 &       0 & $0^+_S$ \\
& & & & & & & & & & & \\
\hline
\end{tabular}
\end{table}

\clearpage

\begin{table}
\centering
\caption[Nucleon resonances]
{Mass spectrum of nucleon resonances in the oblate top model. 
The masses are given in MeV.}
\label{nucleon} 
\vspace{15pt} 
\begin{tabular}{lcccr}
\hline
& & & & \\
Baryon & Status & State & ($v_1,v_2$) & $M_{\mbox{calc}}$ \\
& & & & \\
\hline
& & & & \\
N$( 939)P_{11}$   & **** & $^{2}8_{ 1/2}[56,0^+]$ & (0,0) &  939 \\
N$(1440)P_{11}$   & **** & $^{2}8_{ 1/2}[56,0^+]$ & (1,0) & 1440 \\
N$(1520)D_{13}$   & **** & $^{2}8_{ 3/2}[70,1^-]$ & (0,0) & 1566 \\
N$(1535)S_{11}$   & **** & $^{2}8_{ 1/2}[70,1^-]$ & (0,0) & 1566 \\
N$(1650)S_{11}$   & **** & $^{4}8_{ 1/2}[70,1^-]$ & (0,0) & 1680 \\
N$(1675)D_{15}$   & **** & $^{4}8_{ 5/2}[70,1^-]$ & (0,0) & 1680 \\
N$(1680)F_{15}$   & **** & $^{2}8_{ 5/2}[56,2^+]$ & (0,0) & 1735 \\
N$(1700)D_{13}$   &  *** & $^{4}8_{ 3/2}[70,1^-]$ & (0,0) & 1680 \\
N$(1710)P_{11}$   &  *** & $^{2}8_{ 1/2}[70,0^+]$ & (0,1) & 1710 \\
N$(1720)P_{13}$   & **** & $^{2}8_{ 3/2}[56,2^+]$ & (0,0) & 1735 \\
N$(1900)P_{13}$   &   ** & $^{2}8_{ 3/2}[70,2^+]$ & (0,0) & 1875 \\
N$(1990)F_{17}$   &   ** & $^{4}8_{ 7/2}[70,2^+]$ & (0,0) & 1972 \\
N$(2000)F_{15}$   &   ** & $^{2}8_{ 5/2}[70,2^+]$ & (0,0) & 1875 \\
                  &      & $^{4}8_{ 5/2}[70,2^+]$ & (0,0) & 1972 \\
N$(2080)D_{13}$   &   ** & $^{2}8_{ 3/2}[70,1^-]$ & (1,0) & 1909 \\
                  &      & $^{4}8_{ 3/2}[70,1^-]$ & (1,0) & 2004 \\
N$(2090)S_{11}$   &    * & $^{2}8_{ 1/2}[70,1^-]$ & (1,0) & 1909 \\
                  &      & $^{4}8_{ 1/2}[70,1^-]$ & (1,0) & 2004 \\
N$(2100)P_{11}$   &    * & $^{2}8_{ 1/2}[70,1^+]$ & (0,1) & 1997 \\
                  &      & $^{4}8_{ 1/2}[70,1^+]$ & (0,1) & 2088 \\
N$(2190)G_{17}$   & **** & $^{2}8_{ 7/2}[70,3^-]$ & (0,0) & 2140 \\
N$(2200)D_{15}$   &   ** & $^{2}8_{ 5/2}[70,3^-]$ & (0,0) & 2140 \\
                  &      & $^{4}8_{ 5/2}[70,3^-]$ & (0,0) & 2225 \\
N$(2220)H_{19}$   & **** & $^{2}8_{ 9/2}[56,4^+]$ & (0,0) & 2267 \\
N$(2250)G_{19}$   & **** & $^{4}8_{ 9/2}[70,3^-]$ & (0,0) & 2225 \\
N$(2600)I_{1,11}$ &  *** & $^{2}8_{11/2}[70,5^-]$ & (0,0) & 2590 \\
N$(2700)K_{1,13}$ &   ** & $^{2}8_{13/2}[56,6^+]$ & (0,0) & 2695 \\
& & & & \\
\hline
\end{tabular}
\end{table}

\clearpage

\begin{table}
\centering
\caption[Delta resonances]
{Mass spectrum of delta resonances in the oblate top model. 
The masses are given in MeV.}
\label{delta} 
\vspace{15pt} 
\begin{tabular}{lcccr}
\hline
& & & & \\
Baryon & Status & State & ($v_1,v_2$) & $M_{\mbox{calc}}$ \\
& & & & \\
\hline
& & & & \\
$\Delta(1232)P_{33}$ & **** & $^{4}10_{3/2}[56,0^+]$ & (0,0) & 1232 \\
$\Delta(1600)P_{33}$ &  *** & $^{4}10_{3/2}[56,0^+]$ & (1,0) & 1646 \\
$\Delta(1620)S_{31}$ & **** & $^{2}10_{1/2}[70,1^-]$ & (0,0) & 1649 \\
$\Delta(1700)D_{33}$ & **** & $^{2}10_{3/2}[70,1^-]$ & (0,0) & 1649 \\
$\Delta(1750)P_{31}$ &    * & $^{2}10_{1/2}[70,0^+]$ & (0,1) & 1786 \\
$\Delta(1900)S_{31}$ &  *** & $^{2}10_{1/2}[70,1^-]$ & (1,0) & 1977 \\
$\Delta(1905)F_{35}$ & **** & $^{4}10_{5/2}[56,2^+]$ & (0,0) & 1909 \\ 
$\Delta(1910)P_{31}$ & **** & $^{4}10_{1/2}[56,2^+]$ & (0,0) & 1909 \\ 
$\Delta(1920)P_{33}$ &  *** & $^{4}10_{3/2}[56,2^+]$ & (0,0) & 1909 \\ 
$\Delta(1930)D_{35}$ &  *** & $^{2}10_{5/2}[70,2^-]$ & (0,0) & 1945 \\ 
$\Delta(1940)D_{33}$ &    * & $^{2}10_{3/2}[70,2^-]$ & (0,0) & 1945 \\ 
$\Delta(1950)F_{37}$ & **** & $^{4}10_{7/2}[56,2^+]$ & (0,0) & 1909 \\ 
$\Delta(2000)F_{35}$ &   ** & $^{2}10_{5/2}[70,2^+]$ & (0,0) & 1945 \\
$\Delta(2150)S_{31}$ &    * & $^{4}10_{1/2}[56,1^-]$ & (0,1) & 2029 \\
                     &      & $^{2}10_{1/2}[70,1^-]$ & (0,1) & 2062 \\
$\Delta(2200)G_{37}$ &    * & $^{2}10_{7/2}[70,3^-]$ & (0,0) & 2201 \\
$\Delta(2300)H_{39}$ &   ** & $^{4}10_{9/2}[56,4^+]$ & (0,0) & 2403 \\
                     &      & $^{2}10_{9/2}[70,4^+]$ & (0,0) & 2431 \\
$\Delta(2350)D_{35}$ &    * & $^{4}10_{5/2}[56,3^-]$ & (0,0) & 2170 \\
                     &      & $^{2}10_{5/2}[70,3^-]$ & (0,0) & 2201 \\
$\Delta(2390)F_{37}$ &    * & $^{2}10_{7/2}[70,3^+]$ & (0,0) & 2201 \\
                     &      & $^{4}10_{7/2}[56,4^+]$ & (0,0) & 2403 \\
                     &      & $^{2}10_{7/2}[70,4^+]$ & (0,0) & 2431 \\
$\Delta(2400)G_{39}$ &   ** & $^{2}10_{9/2}[70,4^-]$ & (0,0) & 2431 \\
$\Delta(2420)H_{3,11}$ & **** & $^{4}10_{11/2}[56,4^+]$ & (0,0) & 2403 \\
$\Delta(2750)I_{3,13}$ &   ** & $^{4}10_{13/2}[56,5^-]$ & (0,0) & 2615 \\
                       &      & $^{2}10_{13/2}[70,6^-]$ & (0,0) & 2835 \\
$\Delta(2950)K_{3,15}$ &   ** & $^{4}10_{15/2}[56,6^+]$ & (0,0) & 2811 \\
& & & & \\
\hline
\end{tabular}
\end{table}

\clearpage

\clearpage

\begin{figure}
\centering
\setlength{\unitlength}{1.0pt}
\begin{picture}(360,240)(0,0)
\thinlines
\put (  0,  0) {\line(1,0){360}}
\put (  0,240) {\line(1,0){360}}
\put (  0,  0) {\line(0,1){240}}
\put (360,  0) {\line(0,1){240}}
\thicklines
\put ( 30,180) {\line(1,0){100}}
\put (150,140) {\line(1,0){80}}
\put (150,180) {\line(1,0){80}}
\put (250, 60) {\line(1,0){40}}
\put (250,140) {\line(1,0){40}}
\put (250,180) {\line(1,0){40}}
\put (310, 60) {\line(1,0){20}}
\put (310, 90) {\line(1,0){20}}
\put (310,140) {\line(1,0){20}}
\put (310,170) {\line(1,0){20}}
\put (310,190) {\line(1,0){20}}
\multiput (130,180)(5, 0){5}{\circle*{0.1}}
\multiput (130,180)(2.24,-4.47){9}{\circle*{0.1}}
\multiput (230,180)(5, 0){5}{\circle*{0.1}}
\multiput (230,140)(5, 0){5}{\circle*{0.1}}
\multiput (230,140)(1.21,-4.85){17}{\circle*{0.1}}
\multiput (290,180)(4.47, 2.24){5}{\circle*{0.1}}
\multiput (290,140)(5, 0){5}{\circle*{0.1}}
\multiput (290,140)(2.77, 4.16){8}{\circle*{0.1}}
\multiput (290, 60)(5, 0){5}{\circle*{0.1}}
\multiput (290, 60)(2.77, 4.16){8}{\circle*{0.1}}
\thinlines
\put ( 35,185) {$2^+_S,2^+_M,1^+_A,0^+_S,0^+_M$}
\put (185,185) {$1^+_A$}
\put (155,145) {$2^+_S,2^+_M,0^+_S,0^+_M$}
\put (265,185) {$1^+_A$}
\put (255,145) {$2^+_S,0^+_S$}
\put (255, 65) {$2^+_M,0^+_M$}
\put (315,195) {$1^+_A$}
\put (315,175) {$2^+_S$}
\put (315,145) {$0^+_S$}
\put (315, 95) {$2^+_M$}
\put (315, 65) {$0^+_M$}
\put ( 35, 55) {$U(6) \supset {\cal SU}(3) \otimes SU(2)$}
\put (185, 30) {$\alpha$}
\put (265, 30) {$\kappa'$}
\put (315, 30) {$\kappa$}
\end{picture}
\vspace{15pt}
\caption[The $U(6) \supset {\cal SU}(3) \otimes SU(2)$ limit]
{Splitting of the $n=2$ multiplet in the 
$U(6) \supset {\cal SU}(3) \otimes SU(2)$ limit. 
The masses are calculated using Eq.~(\ref{ener1}) 
by successively adding the terms 
with $\alpha<0$, $\kappa'<0$ and $\kappa>0$.}
\label{ho1}
\end{figure}

\clearpage

\clearpage

\clearpage

\clearpage

\clearpage

\clearpage

\clearpage

\end{document}